\documentclass[%
 reprint,
 showpacs,
 amsmath,amssymb,
 aps,
 prc,
]{revtex4-1}

\usepackage{graphicx}
\usepackage{dcolumn}
\usepackage{bm}

\usepackage{graphicx}	

\usepackage{color}
\usepackage{multirow}
\usepackage{hyperref}
\hypersetup{colorlinks=true, urlcolor=cyan, linkcolor=blue, citecolor=blue}

\usepackage{xspace}	

\newcommand{\ampt}{{\sc ampt}\xspace}
\newcommand{\hijing}{{\sc hijing}\xspace}
\newcommand{\bamps}{{\sc bamps}\xspace}
\newcommand{\vini}{{\sc vini}\xspace}
\newcommand{\mpc}{{\sc mpc}\xspace}
\newcommand{\zpc}{{\sc zpc}\xspace}
\newcommand{\pp}{\mbox{$p$+$p$}\xspace}

\newcommand{\pa}{\mbox{$p$$+$A}\xspace}

\newcommand{\da}{\mbox{$d$$+$A}\xspace}
\newcommand{\nucnuc}{\mbox{A$+$A}\xspace}

\newcommand{\hea}{\mbox{$^3$He$+$A}\xspace}

\newcommand{\ee}{$e^{+}$$e^{-}$\xspace}
\newcommand{\zmass}{$m_{Z}$\xspace}
\newcommand{\zboson}{$Z$\xspace}
\newcommand{\pt}{$p_{T}$\xspace}

\begin{document}

\preprint{APS/123-QED}

\title{Are minimal conditions for collectivity met in \ee collisions? } 

\author{J.L Nagle}
\email{jamie.nagle@colorado.edu}
\author{Ron Belmont}%
\author{Kurt Hill}%
\author{Javier Orjuela Koop}%
\author{Dennis V. Perepelitsa}%
\author{Pengqi Yin}
\affiliation{%
 University of Colorado Boulder, Boulder, CO 80309, USA
}%

\author{Zi-Wei Lin}
\affiliation{East Carolina University, Department of Physics, Greenville, NC 27858, USA}
\affiliation{Key Laboratory of Quarks and Lepton Physics (MOE) and Institute of Particle Physics, Central China Normal University, Wuhan 430079, China}

\author{Darren McGlinchey}
\affiliation{Los Alamos National Laboratory, Los Alamos, NM 87545, USA}%

\date{\today}

\begin{abstract}
Signatures of collective behavior have been measured in highly relativistic \pp collisions, as well as in \pa, \da, and \hea collisions. Numerous particle correlation measurements in these systems have been successfully described by calculations based on viscous hydrodynamic and transport models. These observations raise the question of the minimum necessary conditions for a system to exhibit collectivity. Recently, numerous scientists have raised the question of whether the quarks and gluons generated in \ee collisions may satisfy these minimum conditions. In this paper we explore possible signatures of collectivity, or lack thereof, in \ee collisions utilizing A Multi-Phase Transport (\ampt) framework which comprises melted color strings, parton scattering, hadronization, and hadron re-scattering.
\end{abstract}

 \maketitle


\section{Introduction}
High-energy heavy ion collisions at the Relativistic Heavy Ion Collider (RHIC) and the Large Hadron Collider (LHC) liberate tens of thousands of quarks and gluons that form a medium which is amenable to description by nearly inviscid hydrodynamics---see, for example, Ref.~\cite{Heinz:2013th}. 
Observables such as hadron yields, particle spectra, and multi-particle azimuthal correlations are well described within this framework, supporting the well-established conclusion that the Quark-Gluon Plasma (QGP) behaves as a nearly perfect fluid. However, similar experimental signatures have now been measured in \pp collisions at the LHC~\cite{Aad:2015gqa,Khachatryan:2015lva}, as well as in proton or light ion-nucleus collisions (\pa, \da, \hea) at RHIC~\cite{Nagle:2013lja,Adare:2014keg,Adare:2015ctn,Aidala:2016vgl} and the LHC~\cite{Abelev:2012ola,Aad:2012gla,CMS:2012qk}. In fact, multi-particle azimuthal correlations have now been measured in \pa or \da collisions at a wide range of center-of-mass energies, from 19.6 GeV~\cite{Belmont:2017lse,McGlinchey:2017esf} all the way up to 5.02 TeV~\cite{Khachatryan:2015waa}.    

These results have sparked intense scientific discussion about the minimum size, minimum lifetime, and the minimum number of initial quarks and gluons that are necessary for QGP formation. What does the success of viscous hydrodynamic calculations all the way from \pp to \pa to \nucnuc ~\cite{Weller:2017tsr} tell us about these minimum conditions? In addition to a hydrodynamical picture, alternative proposals exist based on initial-state momentum domains~\cite{Schenke:2017bog,Dusling:2013qoz}. However, the collision geometry dependence of these signatures in \pa, \da, \hea data from RHIC~\cite{Aidala:2016vgl} currently pose a major challenge for these alternatives.   

Within the hydrodynamic framework, there is little evidence for the existence of a minimum system size, or a resulting number of final-state hadrons, below which the successful description of experimental data breaks down~\cite{Koop:2015trj}. The lifetime of the QGP phase is reduced for small systems and for systems with lower initial temperature, but one still finds substantial translation of initial geometry anisotropy into momentum space anisotropies amongst final-state hadrons. However, it has recently been noted that the agreement of viscous hydrodynamics with data does not necessarily imply local equilibration of the QGP, and may rather reflect non-hydrodynamic modes rapidly damping away, even as the system never comes close to  equilibration~\cite{Romatschke:2016hle,Romatschke:2017vte}.   This may explain the so-called ``fast hydrodynamization" (i.e. the early time after which a hydrodynamic description works) and why hydrodynamics works for systems 
where the fundamental assumption of hydrodynamics that the system size be much larger than the nominal mean free path $\bar{R} \gg \lambda$ is violated.  Recent investigations may point to minimum conditions such that these non-hydrodynamic modes do not dominate~\cite{Spalinski:2016fnj}.

In addition to hydrodynamics, alternative frameworks have been proposed for understanding the translation of initial geometry into collective correlations amongst final-state hadrons. These calculations involve microscopic transport models (e.g., \ampt ~\cite{Lin:2004en}, \bamps ~\cite{Xu:2004mz}, \vini ~\cite{Geiger:1991nj}, \mpc ~\cite{Molnar:2000jh}) that trace well-defined individual constituents (quarks and gluons) and generate collectivity via a modest number of scatterings between them~\cite{He:2015hfa,Lin:2015ucn}.   

This discussion of small hadronic collision systems raises the question of whether a high enough parton multiplicity exists in \ee collisions for signatures of collectivity to be clearly observed. The yields and spectra of hadrons have been discussed in terms of thermal models~\cite{Becattini:2008tx,Becattini:1996gy,Becattini:2001fg,Back:2003xk}, though no clear conclusions have been drawn from them.   Thermodynamical string fragmentation has also been studied in comparing \ee and \pp collision results~\cite{Fischer:2016zzs}.  Other types of collisions involving deep inelastic scattering of electrons from protons or nuclei, available for example at a future Electron-Ion Collider, also provide scenarios where the same question could be raised. In this paper, we focus on the simpler \ee case.

The Large Electron-Positron Collider (LEP) operated from 1989-2000, and collided \ee at center-of-mass energies of up to 209 GeV~\cite{Olive:2016xmw}.  When the $e^{+}$ and $e^{-}$ collide, they annihilate and form a virtual photon or \zboson boson. Here we focus on collisions at a center-of-mass energy at the \zboson boson mass. The \zboson boson can decay into a quark and anti-quark pair moving in opposite directions in the \zboson boson's center-of-mass frame. In a simplified picture, a color string extends between the quark and anti-quark as they recede away from each other~\cite{string}.   
This string picture treats all but the highest-energy partons as field lines, which are attracted to each other via the gluon self-interaction, thus forming a narrow flux tube of color field.
The color field in the string provides the necessary energy for new quark-antiquark pairs ($q\overline{q}$) or diquark-antidiquark pairs ($qq~\overline{qq}$) to be emitted along the length of the string.  Ignoring higher-order QCD effects, is it possible that this single color string with parton emissions and subsequent interactions among these partons is sufficient to generate collectivity?

\section{A Multi-Phase Transport}

A Multi-Phase Transport (\ampt) provides a framework for the modeling of relativistic heavy ion collisions~\cite{Lin:2004en}.   \ampt version 2.26 is publicly available~\cite{publicampt} and models these collisions as a succession of distinct stages in time. Initially, the nucleons within the colliding nuclei are spatially distributed via a Monte Carlo Glauber calculation, as implemented within the \hijing model~\cite{Gyulassy:1994ew}. For each event, the impact parameter is randomly chosen and nucleon-nucleon collisions result in the formation of color strings. These color strings then produce quarks and antiquarks via a process referred to as ``string melting".   Quarks and antiquarks emitted from this process are then allowed to interact with each other, as coded within the \zpc parton cascade~\cite{Zhang:1997ej}, with an effective parton-parton cross section provided as an input parameter to the code. Hadronization is implemented via spatial quark coalescence.   
After coalescence, final-state hadronic scattering takes place until the density of hadrons is such that interactions cease to occur. This framework has been reasonably successful at describing a number of features, including particle spectra, yields, and azimuthal correlations, in systems ranging from \pa to \da to \hea to \nucnuc at RHIC and the LHC. For example, see Refs.~\cite{Bzdak:2014dia,Koop:2015wea,Koop:2015trj,Adare:2015cpn,Ma:2016fve}.   

The \ampt model reports the entire history of the partons in both space-time and momentum-energy coordinates, as well as the final distribution of hadrons after final-state scattering. One also has the ability to tune the strength of, and even to fully deactivate, partonic and hadronic scattering.

\section{Modeling $e^{+}e^{-}$ Collisions in \ampt}

We have written a modified version of \ampt where every event is initialized with a single color string, and where the receding quark and antiquark ($u\overline{u}$ or $d\overline{d}$ or $s\overline{s}$) are moving parallel to the longitudinal ($z$) axis.   One can think of this as experimentally rotating into a coordinate system along the two-jet thrust axis. The center-of-mass energy of the string is set in every event to correspond to the \zboson boson mass, \zmass $= 91.18$ GeV$/c^{2}$.   Without further 
modification, the \ampt code then carries out ``string melting", resulting in an initial space-momentum distribution of partons in each event. The partons are then
allowed to interact, then to coalescence into hadrons, and the resulting
hadrons to scatter.  We have set the parton-parton scattering cross section to $\sigma_{parton}$ = 3.7~mb. This value is at the higher end of the range of values typically used in publications to match data both in large and small systems, typically 0.75 - 4.5~mb.

\begin{figure}
\includegraphics[width=0.98\columnwidth]{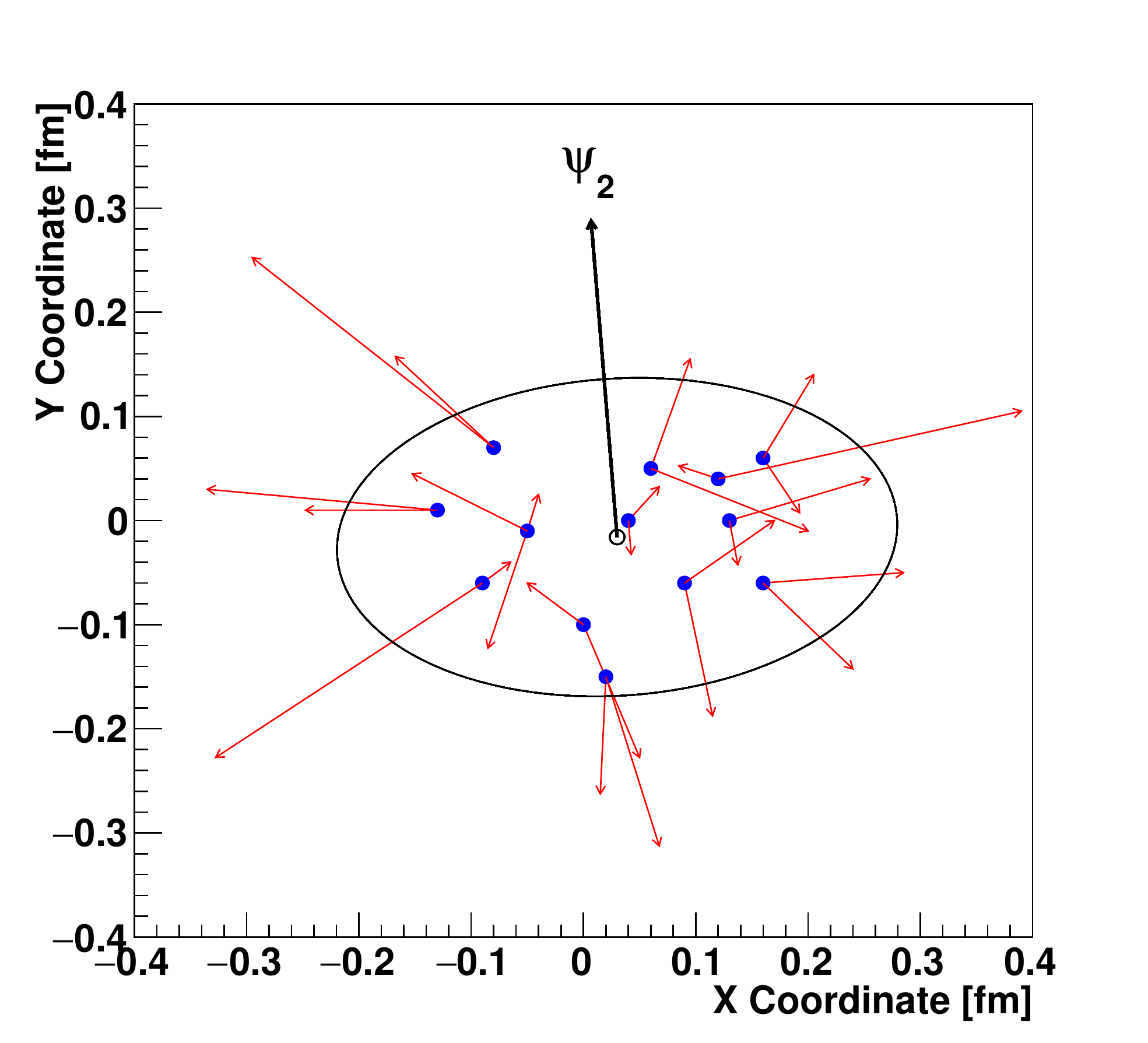}
\caption{Single color-string event with partons in their initial positions shown as blue points, and initial parton momentum vectors shown as red arrows. The center-of-mass coordinate for the set of partons is shown as the black open circle and the spatial eccentricity shown as the ellipse.\label{fig:onestringevent}}
\end{figure}

An example event in its partonic phase is shown in Figure~\ref{fig:onestringevent}.  In this particular event there are approximately 30 partons resulting from string melting.    A peculiarity of \ampt is that all partons are quarks and anti-quarks, with no gluons, and they are constructed as if hadrons were emitted from the string; that is, they are emitted in $q\overline{q}$ pairs (mesons) and $qqq$ triplets (baryons).   A hadron formation time $\tau_{f} = E_{H}/m_{T,H}^{2}$ sets the time for these partons to be available for scattering where $E_{H}$ and $m_{T,H}$ are the energy and transverse mass of the hadron prior to melting into partons~\cite{Lin:2004en}.  As noted explicitly in Ref.~\cite{Lin:2004en}, the "typical string fragmentation time of about 1 fm/c is not applied to the melting of strings as the fragmentation process involved here is considered just as an intermediate step in modeling parton production from the energy
field of the strings in an environment of high energy density."
We find typical formation time values of $\tau_{f} = 0.1-0.3$~fm/c, with a long tail to larger times.
We only display and analyze partons with formation time $\tau_{f} < 3$ fm/c and evaluate their spatial coordinates at the parton's individual formation time.

The blue markers indicate the initial quark positions and the red arrows, their initial momentum vectors (with the vector length proportional to the momentum magnitude). The spatial eccentricity is often characterized via $\varepsilon_{2}$, defined as 
\begin{equation}
\varepsilon_{2} = {{\sqrt{\left< r^{2}\cos(2\phi) \right>^{2} + \left< r^{2}\sin(2\phi) \right>^{2}}\over{\left< r^{2} \right>}}}
\end{equation}
where the averages are over all initial parton transverse coordinates ($r_{i}, \phi_{i}$) relative to their center-of-mass coordinate. The vector $\psi_{2}$ is defined by the short axis of the ellipse.  The black open circle corresponds to the center-of-mass coordinate of the total set of partons.   The ellipse represents their spatial orientation and eccentricity, and the angle $\psi_{2}$ shows the orientation of the eccentricity. For this specific event the initial parton eccentricity is $\varepsilon_{2} = 0.34$.   

In \ampt, the created partons are distributed around the center of the string in the transverse plane with an average radius of approximately 0.1-0.2~fm.   The eccentricity thus varies event-by-event with a rather broad distribution in $\varepsilon_{2}$ from 0 to 1, since there is no intrinsic geometry and just fluctuations of the initial parton coordinates.  The partons are ``born" after some initial momentum-dependent formation time with a strong correlation between their radial position and momentum vector, i.e. moving predominantly radially outward.   

The measured \ee mean charged particle multiplicity at 91 GeV is $\approx$ 21~\cite{Olive:2016xmw,Abbiendi:2004pr}.   We have tuned the \ampt input value \texttt{PARJ(41)}$= 2.5$, which corresponds to parameter $b$ of the {\sc{lund}} symmetric splitting function 
\begin{equation}
f(z) \propto z^{-1}(1-z)^{a}\exp(-bm_{T}^{2}/z)
\end{equation}
where $z$ is the momentum fraction of the produced particle with respect to the fragmenting string and $m_{T}$ is the transverse mass of the produced particle.
We then achieve a mean multiplicity of initial partons $\left<N_{parton} \right> \approx 43$ and a mean charged particle multiplicity of $\left<N_{ch} \right> \approx 21$.   It is notable that for \ampt to describe \pp and \nucnuc multiplicities at different collision energies, these {\sc{lund}} parameters are tuned~\cite{Lin:2014tya}.

\section{Single String Results}

\begin{figure*}[]
\includegraphics[width=0.98\textwidth]{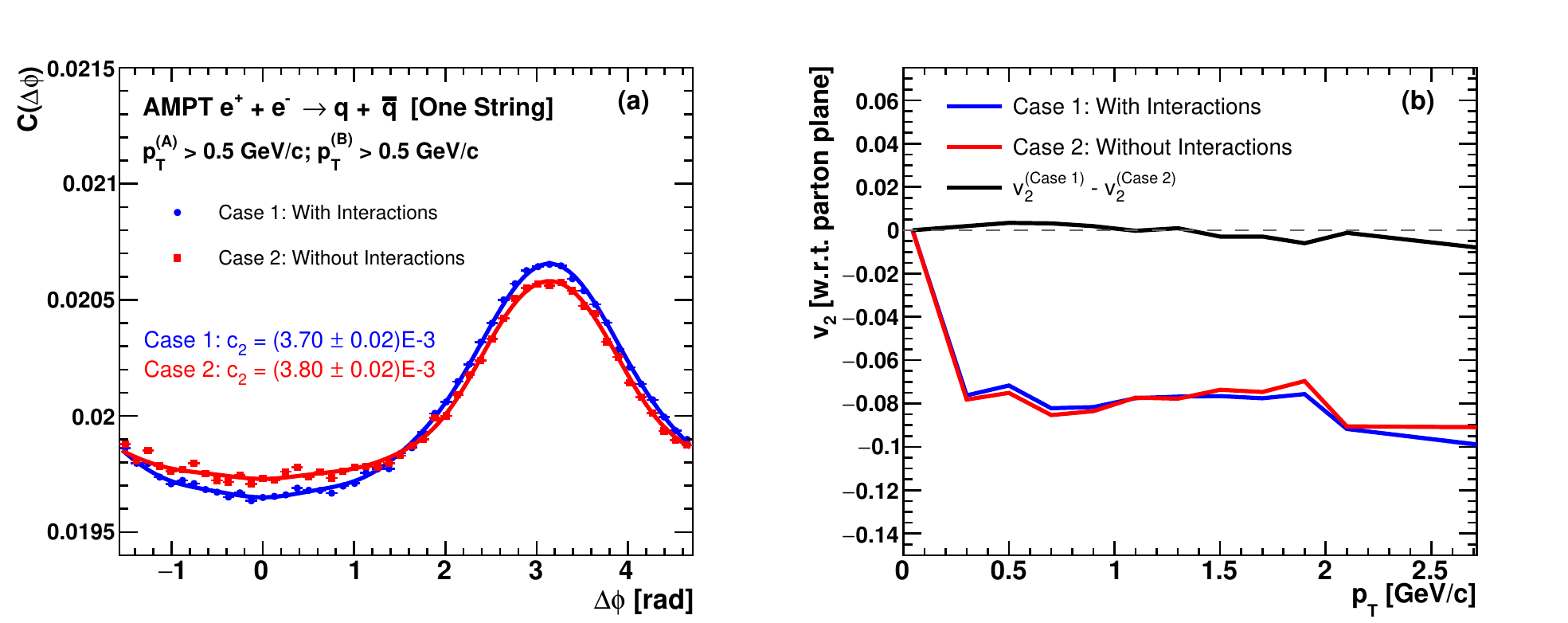}
\caption{\ampt events with a single color string modeling \ee $\rightarrow$ \zboson $\rightarrow q\overline{q}$.   
(Left) Long-range two-particle correlations $|\Delta\eta| > 2.0$ for hadrons with $p_{T} > 0.5$~GeV/$c$, with and without final-state interactions.   Fourier fits are also shown as lines and the $c_{2}$ coefficients displayed.  
(Right) Azimuthal anisotropy ($v_{2}$) calculated with respect to the initial parton plane with and without final-state interactions, and then the net difference thus isolating the effects due to final-state interactions (black curve).
\label{fig:onestringresults}}
\end{figure*}

In order to explore the single color string case, we have run 300 million such events. It is notable that only 17.9\% of all partons resulting from string melting undergo one or more partonic scatterings.  For comparison, we have run another 300 million events where all final-state partonic and hadronic interactions were turned off. The latter gives us a baseline where only initial-state momentum correlations exist and, by definition, there is no collectivity in the final-state.

We have constructed long-range two-particle correlations by taking all pairs of final-state hadrons with $p_{T} > 0.5$~GeV/$c$ and calculating their relative angle $\Delta \phi$ with an imposed pseudorapidity gap $|\Delta\eta| < 2.0$. Note again that pseudorapidity is defined by treating the string axis as the longitudinal axis.  Figure~\ref{fig:onestringresults} (left panel) shows the long-range two-particle correlation for hadrons with $p_{T} \ge 0.5$ GeV/$c$ in the one string case with (blue) and without (red) final-state interactions.    
Also shown are Fourier component characterizations for both cases and quoted the second Fourier component ($c_{2} = \left< \cos(2\Delta\phi) \right>$).
The two distributions are nearly identical indicating that for this system, final-state interactions play a negligible role in correlated particle yields across a rapidity gap.

The single color string case for modeling \ee collisions does incorporate a modest number of final-state parton-parton scatterings. However, these scatterings have a very small effect on the final rapidity-separated collectivity signature.  This may not be surprising for the long range (large $\Delta\eta$) two-particle correlations since the initial geometric eccentricity is driven in this case solely by fluctuations in the spatial coordinates of the partons and should be mostly uncorrelated for different rapidity slices.    

Another test for the impact of partonic scattering on collectivity, that does not require long-range rapidity correlations, is to check for azimuthal anisotropy ($v_{2}$) with respect to the initial geometry orientation (i.e. $\psi_{2}$) as determined from the partons that emerge from string melting--- as shown in Figure~\ref{fig:onestringevent}, for example.   The elliptic azimuthal anisotropy $v_{2}$ is defined by
\begin{equation}
v_{2} = \left< \cos(2(\phi-\psi_{2})) \right>.
\end{equation}
Whereas here we use early stage partons to define the event geometry, previous \ampt studies of small system collectivity have calculated $v_2$ relative to the orientation defined by participant nucleons~\cite{Koop:2015trj,Koop:2015wea}. Figure~\ref{fig:onestringresults} (right panel) shows $v_{2} (p_{T})$ with respect to the parton plane for final-state hadrons. The results with interactions (blue) and without interactions (red) both yield large negative values for $v_{2}$.   

The explanation can be understood by examining Figure~\ref{fig:onestringevent}.
In this event there is a fluctuation to have more partons along the major axis of the ellipse, oriented along the $y=0$ line. The partons are born with large radial outward momentum (i.e. they have a strong radial position-momentum correlation), and so there are more partons initially moving to the left and to the right, as opposed to up and down. Since the $\psi_{2}$ vector is nearly up, the particle distribution with no interactions already has a large negative $v_{2}$. The black curve in Figure~\ref{fig:onestringresults} shows the net effect of interactions on the $v_{2}$, which is very modest, less than a 1\% effect.

\section{Two String Results}

\begin{figure}
\includegraphics[width=0.98\columnwidth]{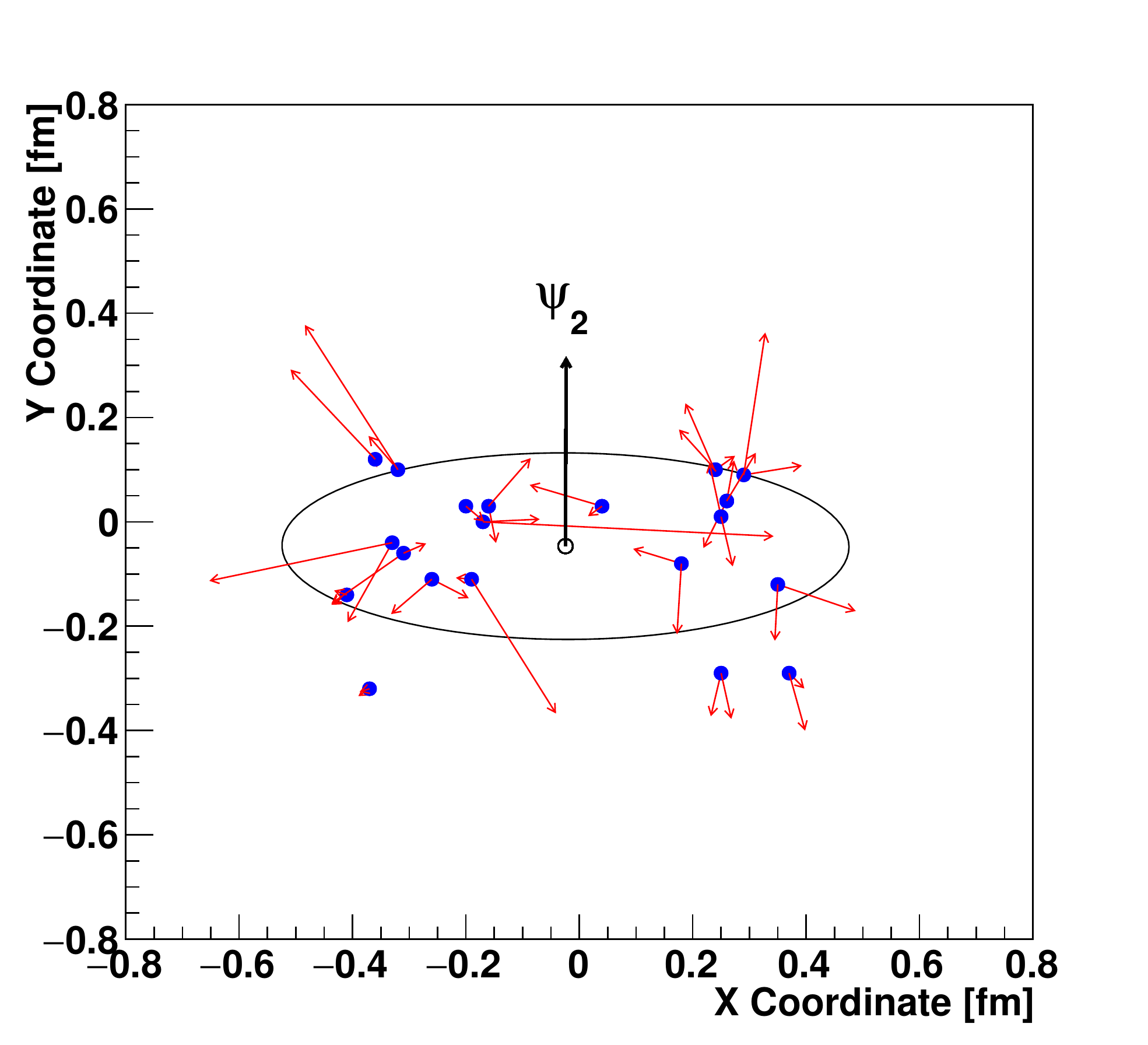}
\caption{Two color-string event with initial parton positions shown as blue points, and initial parton momentum vectors shown as red arrows.   The center-of-mass coordinate for the set of partons is shown as the black open point and the spatial eccentricity shown as the drawn ellipse.\label{fig:twostringevent}}
\end{figure}

In order to further explore the minimum conditions for collectivity, we have extended these calculations to the special case of two color strings separated by 0.5~fm in the transverse plane.   Each color string is identical to those considered above with the string oriented perfectly along the longitudinal direction, except each has an energy corresponding to the half the \zboson boson mass. A single event display is shown in Figure~\ref{fig:twostringevent}.  This event has 40 partons initiated from string melting and an initial parton spatial eccentricity of $\varepsilon_{2} = 0.64$.

\begin{figure*}[ht!]
\includegraphics[width=0.98\textwidth]{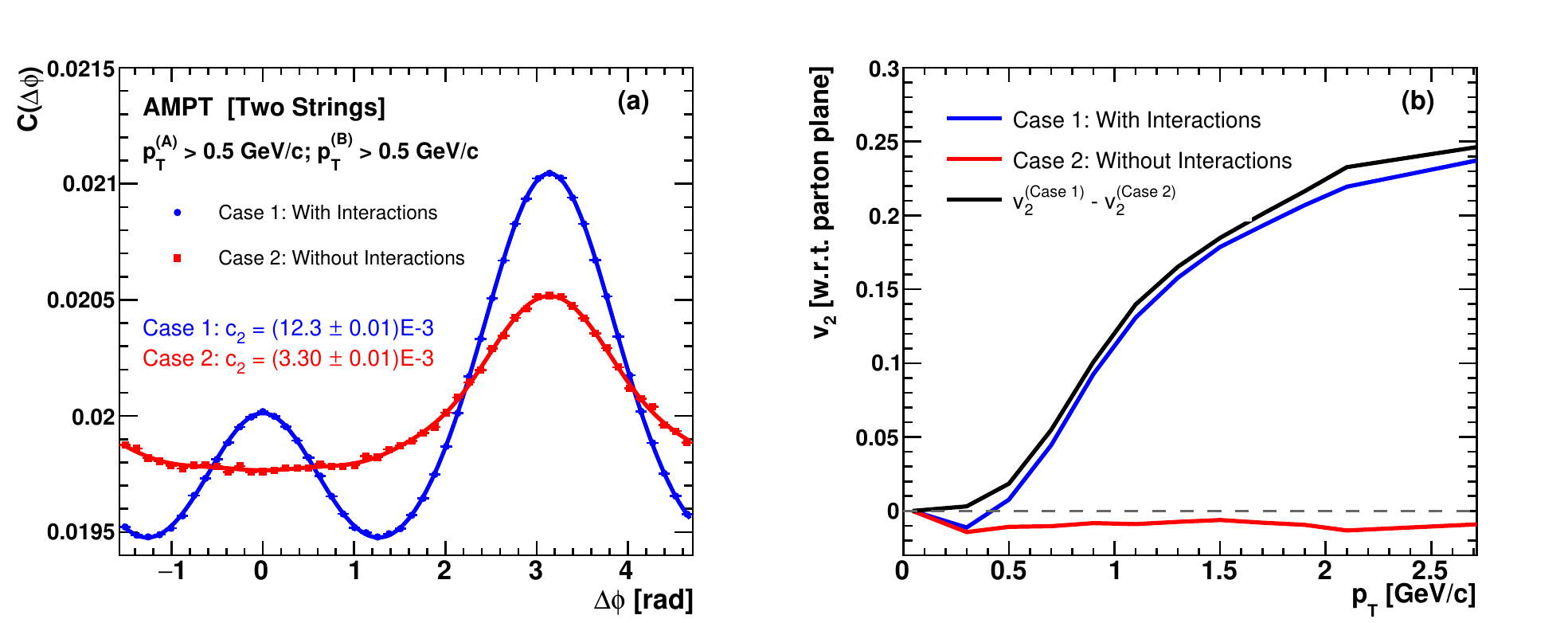}
\caption{
\ampt events with two color strings.   
(Left) Long-range two-particle correlations $|\Delta\eta| > 2.0$ for hadrons with \pt $> 0.5$ GeV/$c$, with and without final-state interactions.   Fourier fits are also shown as lines and the $c_{2}$ coefficients displayed.  
(Right) Azimuthal anisotropy ($v_{2}$) calculated with respect to the initial parton plane with and without final-state interactions, and then the net difference thus isolating the effects due to final-state interactions (black curve).
\label{fig:twostringresults}}
\end{figure*}

There are two key differences from the single color string case. First, the number of parton scatterings is substantially increased (even when controlling for the higher total multiplicity of partons). In this case 40.3\% of all partons suffer one or more scatterings.    It is notable that this number is already quite similar to the percentage of partons that scatter in central \da collisions at center-of-mass energies from 19.6-200 GeV~\cite{Koop:2015wea}.  In this two string case, there are partons from the left string moving right and partons from the right string moving left (i.e. towards each other).   Second, there is now a long-range correlation in the geometry of the initial parton coordinates. The $\psi_{2}$ axis is predominantly perpendicular to the axis connecting the two strings (the $x-$axis in Figure~\ref{fig:twostringevent}), and this is true in all rapidity slices modulo fluctuations.

Figure~\ref{fig:twostringresults} shows the results with (blue) and without (red) final-state partonic and hadronic interactions for the long-range two-particle correlations (left panel) and the $v_{2}$ with respect to the parton plane (right panel).   The results are quite striking in that there is now a visible near-side ridge (a local maximum near $\Delta\phi = 0$) in the long-range correlation that is only seen when interactions are turned on.   We again show a Fourier component characterization and quote the $c_{2}$ coefficient, and find a much larger coefficient in the case with interactions.

Even more striking is the very large net $v_{2}$ as a function of \pt seen with respect to the parton plane (shown as the black curve). In fact, this $v_{2}$ result (again, having accounted for the initial negative $v_{2}$ from momentum correlations alone), is quite similar in magnitude and \pt trend to that observed in \pp collisions and various light ion-nucleus collision results.

We also ran this two string configuration with just partonic scattering, i.e. hadronic rescattering turned off, and the $v_{2}$ results are nearly identical.    Thus, the dominant contribution arises from parton scattering alone.

\section{Two String Energy Dependence}

While the two string scenario does not correspond to an exact physical interaction system, it is interesting to explore the energy dependence of the correlation and flow observables.   To that end, we have run \ampt with the identical configuration of two strings as detailed above at different total collision energies (184, 91, 60, 45, 30, 10, and 4.5~GeV).  For each energy, the
resulting total number of partons produced, the $dN_{ch}/d\eta$ within the window $|\eta|<2.0$, and the percentage of partons that have at least one scattering are shown in Table~\ref{tab:twostringenergy}.  There is a substantial decrease in the number of partons produced at lower energies, as expected; however, since the extension of the strings in rapidity is also being reduced, the probability of scattering has a much weaker energy dependence.

\begin{table}[h]
	\centering
	\caption{\label{tab:twostringenergy} 
    Summary of two string results as a function of the total available energy.    Columns include the total number of partons over all phase space, $dN_{ch}/d\eta$ within the window $|\eta|<2.0$, and the percentage of partons that have at least one scattering, i.e. $N_{scatter}>0$.}
    \bigskip
	\begin{tabular}{c|c|c|c}
	\hline\hline
    Energy & $N_{partons}$ & $dN_{ch}/d\eta$ & \% of Partons  \\ 
    (GeV) & - & ($|\eta|<2$) & w/ $N_{scatter}>0$ \\ \hline
    184 & 95 & 11.6  & 40.4\% \\ 
    91  & 75 & 11.2 & 40.3\% \\
    60  & 63 & 10.5  & 39.2\% \\
    45  & 55 & 9.7 & 38.7\% \\
    30  & 44 & 8.3  & 37.6\% \\
    10  & 19 & 6.5 & 24.6\% \\
    4.5 & 9 & 3.7 & 11.3\% \\
	\hline\hline
	\end{tabular}
\end{table}

\begin{figure*}
\includegraphics[width=0.98\textwidth]{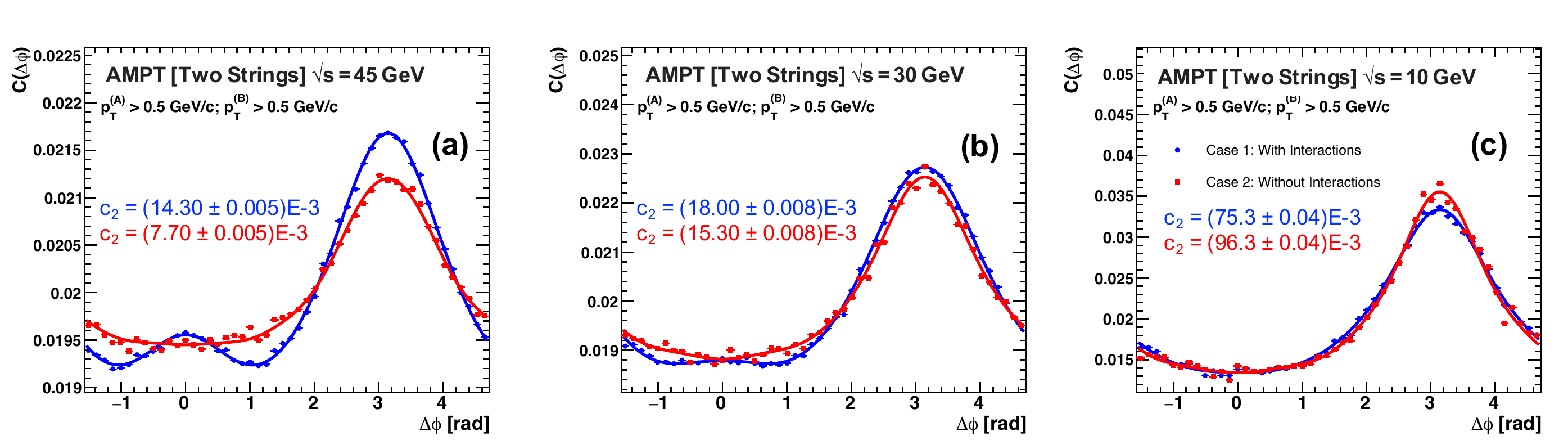}
\caption{
\ampt events with two color strings at total collision energies of 45, 30, and 10 GeV.   Shown are the long-range two-particle correlations $|\Delta\eta| > 2.0$ for hadrons with \pt $> 0.5$ GeV/$c$, with and without final-state interactions.   Fourier fits are also shown as lines and the $c_{2}$ coefficients displayed.
\label{fig:twostringenergyscan}}
\end{figure*}

Figure~\ref{fig:twostringenergyscan} shows the two-particle azimuthal correlations with a pseudorapidity gap $|\Delta \eta| > 2.0$ for total energies of 45, 30, and 10 GeV.    The ``ridge" feature is visible for the higher energies all the way down to 45 GeV.   There is a hint of the "ridge" feature at 30 GeV, and then it disappears for lower energies.    The very large correlation peak at $\Delta\phi = \pi$ from momentum conservation becomes dominant at the lowest energies --- note the change in the vertical scale between the panels.   

\begin{figure*}
\includegraphics[width=0.7\textwidth]{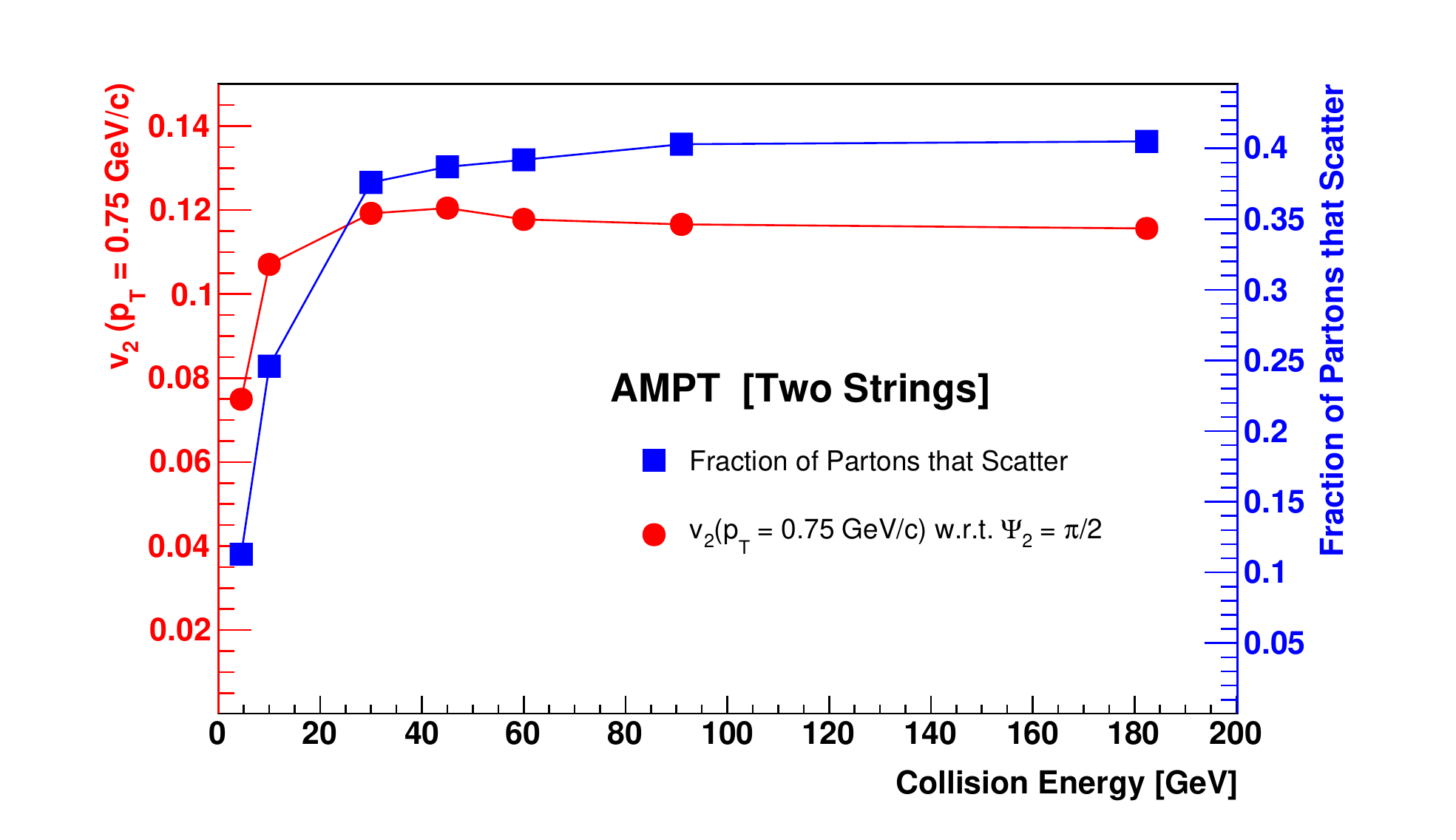}
\caption{
\ampt calculated $v_{2}$ at $p_{T}=0.75$~GeV/c with respect to the true geometry in the two color string configuration as a function of the total energy.  Also shown superimposed is the fraction of partons that suffer at least one scattering.
\label{fig:energyscan}}
\end{figure*}

In addition, we have calculated the $v_{2}$ relative to the true geometry of the two string configuration.   In the previous section, we determined $\Psi_{2}$ event-by-event using the initial parton coordinates.  However, in this case to avoid the fact that $\Psi_{2}$ will have more significant fluctuations at lower energies due to the smaller number of partons in the event, we have simply set $\Psi_{2}$ to be along the axis perpendicular to the line connecting the two strings.   Figure~\ref{fig:energyscan} shows the $v_{2}$ with respect to the true geometry for hadrons with $p_{T} = 0.75$~GeV/c as a function of the two string total energy.   Also superimposed on the same figure is the fraction of partons that undergo at least one scattering.

The signal for the ``ridge" does disappear below some total energy, but that can be attributed in part to the total domination of the momentum conservation correlation as the total number of partons and final state hadrons decreases.   The real effective anisotropy relative to the true geometry, which is not experimentally observable, in fact has very little energy dependence until one goes below a total energy of approximately 10 GeV.  

\section{Summary}

Experimental results from high-energy \pp and light ion-nucleus collisions have prompted the question, what are the minimal conditions for collectivity?  We explore this question for the case of \ee collisions utilizing the \ampt framework and a single color string.    The results indicate only a modest number of parton-parton scatterings and no observable collectivity signal.
However, a simple extension to two color strings predicts finite long-range two-particle correlations (i.e. the ridge) and a strong $v_{2}$ with respect to the initial parton geometry. 
Studying the energy dependence of the signal in the two string configuration reveals a rather robust anisotropy relative to the true geometry even at the lowest energies, though in a regime where extracting the signal via experimentally measured correlations is much more challenging.
These results imply that in small collisions systems over a range
of energies, a minimum of two strings is sufficient to generate collectivity signals.
The question of whether additional mechanisms, such as higher-order effects, could generate such conditions in \ee collisions, or whether they may be present in electron-ion collisions via a picture of a color dipole scattering with the nucleus, remain to be explored.

\section{Acknowledgements}

We acknowledge useful discussions with Peter Steinberg, Jan Fiete Grosse-Oetringhaus, and Bill Zajc.   JLN, RB, KH, JOK, PY acknowledge funding from the Division of Nuclear Physics of the U.S. Department of Energy under Grant No. DE-FG02-00ER41152.  ZWL acknowledges funding from the National Natural Science Foundation of China grant No. 11628508.  DM acknowledges support from the Department of Energy, Office of Science, Nuclear Physics Program.

\bibliographystyle{apsrev4-1}
\bibliography{main}

\end{document}